\documentclass[conference]{IEEEtran}

\usepackage{cite}
\usepackage{amsmath,amssymb,amsfonts}
\usepackage{algorithmic}
\usepackage{graphicx}
\usepackage{textcomp}
\usepackage{xcolor}
\usepackage{booktabs}
\usepackage{hyperref}
\usepackage{cleveref}

\hypersetup{
    colorlinks=true,
    linkcolor=blue,
    filecolor=magenta,      
    urlcolor=cyan,
    citecolor=blue,
}

\def\BibTeX{{\rm B\kern-.05em{\sc i\kern-.025em b}\kern-.08em
    T\kern-.1667em\lower.7ex\hbox{E}\kern-.125emX}}
    
\begin{document}

\title{Comprehensive Evaluation of Large Language Models on Software Engineering Tasks: A Multi-Task Benchmark}

\author{\IEEEauthorblockN{Go Frendi Gunawan\IEEEauthorrefmark{1}, Mukhlis Amien\IEEEauthorrefmark{2}}
\IEEEauthorblockA{\IEEEauthorrefmark{1}gofrendiasgard@gmail.com,
\IEEEauthorrefmark{2}amien@ubhinus.ac.id}}

\maketitle

\begin{abstract}
Large Language Models (LLMs) have demonstrated remarkable capabilities in software engineering, yet comprehensive benchmarks covering diverse SE activities remain limited. We present a multi-task evaluation of 11 state-of-the-art LLMs across five representative software engineering tasks: bug fixing, feature development, code refactoring, technical copywriting, and research synthesis. Our automated verification framework measures both output quality and completion efficiency. Key findings reveal that (1) models achieving identical perfect scores exhibit \textbf{22$\times$ variation in completion time}, \textbf{49$\times$ variation in tool efficiency}, and \textbf{53$\times$ variation in estimated cost}; (2) tool usage frequency shows no correlation with success ($r=0.077$, $p=0.575$)---one model used 917 tool calls while another solved the same task with 3 calls; (3) we identify two distinct inefficiency patterns: \textit{loop inefficiency} (repetitive tool sequences) and \textit{inference inefficiency} (slow token generation); and (4) coding tasks achieve 100\% success while research tasks present greater challenges (90.9\%). These results provide evidence-based guidance for practitioners selecting LLMs based on task requirements, speed constraints, and budget considerations. We release all experimental data, verification scripts, and analysis code for full reproducibility.
\end{abstract}


\section{Introduction}
\label{sec:intro}

Large Language Models (LLMs) have demonstrated remarkable capabilities in software engineering tasks, from code generation and bug fixing to documentation and architectural design \cite{chen2021evaluating, jimenez2023swe}. As these models become increasingly integrated into development workflows, practitioners face a critical question: which model should they choose for specific tasks?

\subsection{Motivation}

The landscape of LLMs for code has expanded rapidly. OpenAI's GPT-4 and its successors have set benchmarks in coding proficiency \cite{openai2024gpt4}. Google's Gemini series offers competitive performance with different architectural approaches \cite{gemini2024}. Open-weight alternatives like Deepseek, GLM, and Qwen provide options for organizations with data privacy or cost constraints \cite{deepseek2024, glm2024, qwen2024}.

However, practitioners face several challenges when selecting models:

\begin{enumerate}
    \item \textbf{Fragmented benchmarks:} Existing evaluations focus on isolated capabilities (e.g., function-level code generation \cite{chen2021evaluating} or bug fixing \cite{jimenez2023swe}) rather than comprehensive software engineering workflows.
    
    \item \textbf{Task-specific studies:} Most evaluations test single task types, leaving unclear how models generalize across diverse activities like refactoring, documentation, and research.
    
    \item \textbf{Efficiency metrics:} Prior work often emphasizes accuracy alone, ignoring practical constraints like completion time and API costs.
    
    \item \textbf{Tool integration:} Modern LLMs operate through agent frameworks with tool use, yet few evaluations measure tool efficiency alongside output quality.
\end{enumerate}

\subsection{Research Questions}

This study addresses these gaps through systematic evaluation of 11 state-of-the-art LLMs across five representative software engineering tasks. We investigate:

\begin{description}
    \item[RQ1] How do current LLMs rank in overall performance across diverse software engineering tasks?
    \item[RQ2] Which models excel at specific task types (coding, writing, research)?
    \item[RQ3] What is the relationship between tool usage frequency and task success?
    \item[RQ4] How do completion time and accuracy correlate across models?
    \item[RQ5] What cost-performance tradeoffs exist across model tiers?
\end{description}

\subsection{Contributions}

Our work makes the following contributions:

\begin{enumerate}
    \item \textbf{Multi-task benchmark:} We present a comprehensive evaluation covering bug fixing, feature development, refactoring, technical writing, and research synthesis with automated verification.

    \item \textbf{Broad model comparison:} We evaluate 11 models from 4 provider categories (OpenAI, Google, Deepseek, Open-weight), enabling cross-provider analysis.

    \item \textbf{Efficiency analysis:} We introduce efficiency metrics (Tool Efficiency Ratio, Time Efficiency Ratio, Cost Estimation) revealing 22--53$\times$ variance among models achieving identical scores.

    \item \textbf{Inefficiency taxonomy:} We identify and categorize two distinct inefficiency patterns (loop vs inference), providing actionable insights for framework optimization.

    \item \textbf{Tool usage insights:} We analyze the relationship between tool invocation patterns and success, finding that more tools do not guarantee better results---917 vs 3 tool calls for identical outcomes.

    \item \textbf{Public dataset:} We release all experimental data, verification scripts, and results to enable reproducibility and future research \cite{our_dataset}.
\end{enumerate}

\subsection{Key Findings}

Our analysis reveals several surprising findings:

\begin{itemize}
    \item Four models (GPT-5.1, Gemini-3 Pro, Deepseek-Chat, GLM-4.7) achieved perfect scores, but exhibited dramatic efficiency variance: \textbf{22$\times$} in completion time (33s vs 732s), \textbf{49$\times$} in tool calls (3.8 vs 188 avg), and \textbf{53$\times$} in estimated cost.

    \item No correlation exists between tool usage count and success (Pearson $r=0.077$, $p=0.575$). GPT-5.1 solved bug-fix in 18.8s with 3 tools; Gemini-3 Flash took 625s with 917 tools---achieving identical EXCELLENT scores.

    \item We identify two distinct inefficiency patterns: (a) \textit{loop inefficiency}, where agents repeat tool sequences without recognizing failure, and (b) \textit{inference inefficiency}, where models generate correct solutions slowly.

    \item Research tasks were most challenging (90.9\% success), while coding tasks achieved 100\% success across all models.

    \item OpenAI models were consistently fastest (avg 54s) without sacrificing quality (9.33 avg score), offering the best speed-quality tradeoff.
\end{itemize}

\subsection{Paper Organization}

The remainder of this paper is organized as follows:
Section~\ref{sec:related} reviews related work in LLM evaluation.
Section~\ref{sec:methodology} describes our task design, model selection, and evaluation framework.
Section~\ref{sec:results} presents quantitative findings and statistical analyses.
Section~\ref{sec:discussion} discusses implications and limitations.
Section~\ref{sec:conclusion} concludes with future directions.

\section{Related Work}
\label{sec:related}

\subsection{Code Generation Benchmarks}

Early benchmarks for code-generating LLMs focused on isolated programming problems. HumanEval \cite{chen2021evaluating} introduced 164 hand-written Python problems with unit test verification, establishing the pass@k metric. Most Basic Python Programming (MBPP) \cite{austin2021program} expanded this to 974 crowd-sourced problems covering diverse programming concepts.

These benchmarks established foundational capabilities but had limitations: (1) function-level scope ignores real-world complexity, (2) no interaction with existing codebases, and (3) limited to algorithmic rather than engineering tasks.

\subsection{Software Engineering Benchmarks}

Recent work addresses these limitations through repository-level evaluation. SWE-bench \cite{jimenez2023swe} presents 2,294 real GitHub issues from popular Python repositories, testing models on actual bug reports and feature requests. While more realistic, SWE-bench has high variance in issue quality and difficulty.

ClassEval \cite{classexplore} focuses on class-level code generation, requiring models to implement multiple interacting methods. RepoBench \cite{repobench} evaluates repository-level code completion with long-context understanding.

Our work complements these by focusing on \textit{controlled, reproducible tasks} with objective verification, enabling precise model comparison.

\subsection{Multi-Task Evaluation}

Prior multi-task studies examined LLM capabilities across domains but not specifically for SE. Hendrycks et al. \cite{hendrycks2021measuring} evaluated coding alongside math and reasoning. Fu et al. \cite{fu2023gptscore} compared GPT models on various NLP tasks.

For SE specifically, "Evaluating LLM-Guided Software Programming" \cite{evaluating2024llm} compared GPT-3.5, GPT-4, and CodeLlama on five SE tasks. Our work expands this to 11 models, adds automated verification, and measures efficiency metrics.

\subsection{Tool-Augmented Evaluation}

Modern LLMs operate through agents with tool use. Works like ToolBench \cite{qinetal2023toolllm} evaluate tool learning, but focus on general tool use rather than SE-specific workflows. Our evaluation framework uses a consistent tool environment across all models, enabling fair comparison of tool efficiency.

\subsection{Efficiency and Cost Analysis}

Recent work recognizes the importance of efficiency. "Efficiency Benchmarking" \cite{efficiency2024} measured inference costs but focused on throughput rather than task completion time. Our work integrates time and quality metrics, revealing that faster models can achieve equal or better results.

\subsection{Research Gap}

Our work fills the gap in:\\
\textbf{(1) Task diversity:} Covering coding, writing, and research synthesis.\\
\textbf{(2) Scale:} 11 models with automated, reproducible verification.\\
\textbf{(3) Efficiency metrics:} Novel Tool Efficiency Ratio and cost estimation, revealing 22--53$\times$ variance.\\
\textbf{(4) Tool analysis:} Quantifying the relationship between tool use and success.\\
\textbf{(5) Anomaly taxonomy:} Identifying loop inefficiency and framework incompatibility patterns.

\section{Methodology}
\label{sec:methodology}

This section describes our experimental design, including task categories, model selection, evaluation framework, and metrics.

\subsection{Task Categories}
\label{sec:tasks}

We designed five representative software engineering tasks that cover different aspects of developer workflows. Each task represents a common activity in professional software development.

\subsubsection{Bug Fixing (bug-fix)}
Participants were presented with \texttt{inventory\_system.py}, a Python script simulating an inventory management system with a concurrency bug. The bug caused race conditions when multiple threads attempted to purchase items simultaneously, resulting in negative inventory counts. Models were required to:
\begin{itemize}
    \item Identify the race condition in multi-threaded code
    \item Implement appropriate synchronization mechanisms
    \item Ensure data integrity under concurrent access
    \item Verify the fix with test scenarios
\end{itemize}

\textit{Success Criteria:} Implementation of proper concurrency control (e.g., locks) and verification that final inventory remains non-negative.

\subsubsection{Feature Implementation (feature)}
Models were given a partial FastAPI Todo application (\texttt{todo\_app.py}) with only a GET endpoint implemented. The task required completing the CRUD operations:
\begin{itemize}
    \item POST /todos -- Create new todo items
    \item PUT /todos/\{id\} -- Update existing items
    \item DELETE /todos/\{id\} -- Delete items
\end{itemize}

Additional requirements included proper HTTP status codes (201 for create, 404 for non-existent items) and auto-incrementing IDs.

\textit{Success Criteria:} All endpoints functional with correct HTTP semantics.

\subsubsection{Code Refactoring (refactor)}
The refactoring task presented \texttt{etl.py}, a monolithic script performing Extract-Transform-Load operations with hardcoded configuration and fragile string parsing. Models were required to:
\begin{itemize}
    \item Separate concerns into distinct ETL phases
    \item Decouple configuration from logic
    \item Implement robust parsing using regular expressions
    \item Add type hints and documentation
    \item Maintain identical output (report.html)
\end{itemize}

\textit{Success Criteria:} Modular architecture with ETL pattern, configuration separation, type hints, docstrings, and functional equivalence.

\subsubsection{Technical Copywriting (copywriting)}
Models were tasked with creating a launch announcement blog post for ``Zrb-Flow,'' a fictional DevOps automation tool. Requirements included:
\begin{itemize}
    \item Mentioning key features: AI, automation, CLI, Docker, K8s
    \item Highlighting ``Self-Healing Pipelines'' capability
    \item Technical but engaging tone
    \item Call to action for installation
    \item Proper Markdown formatting
\end{itemize}

\textit{Success Criteria:} Complete coverage of required keywords, proper formatting, and compelling narrative.

\subsubsection{Technical Research (research)}
Models conducted research on Solid State Batteries (late 2024/2025), producing a comprehensive report covering:
\begin{itemize}
    \item Commercial timeline and automotive applications
    \item Key industry players and companies
    \item Remaining technical challenges
    \item Proper citations and references
\end{itemize}

\textit{Success Criteria:} Substantial content (200+ words), coverage of all three required aspects, and inclusion of references.

\subsection{Model Selection}
\label{sec:models}

We evaluated 11 state-of-the-art Large Language Models from four major provider categories, as shown in Table~\ref{tab:models}.

\begin{table}[htbp]
\centering
\caption{Evaluated Language Models}
\label{tab:models}
\begin{tabular}{llll}
\toprule
\textbf{Provider} & \textbf{Model} & \textbf{Version/Date} & \textbf{Access} \\
\midrule
OpenAI & GPT-4o & 2024 & API \\
& GPT-5.1 & 2025 & API \\
& GPT-5.2 & 2025 & API \\
\midrule
Google & Gemini 2.5 Flash & 2025 & API \\
& Gemini 2.5 Pro & 2025 & API \\
& Gemini 3 Flash & 2025 (preview) & API \\
& Gemini 3 Pro & 2025 (preview) & API \\
\midrule
Deepseek & Deepseek-Chat & 2024 & API \\
\midrule
Open/Ollama & GLM-4.7 & 2025 & Cloud \\
& Kimi-K2.5 & 2025 & Cloud \\
& Qwen3-VL & 2025 & Cloud \\
\bottomrule
\end{tabular}
\end{table}

Model selection criteria included:\\
\textbf{(1) State-of-the-art performance:} All models represent current best-in-class for code generation.\\
\textbf{(2) Diversity of architectures:} We included both proprietary (OpenAI, Google) and open-weight models (GLM, Kimi, Qwen).\\
\textbf{(3) Availability:} Models accessible via API or cloud hosting at time of experimentation.

\subsection{Evaluation Framework}
\label{sec:evaluation}

Each model-task combination was executed in an isolated environment with the following components:

\subsubsection{Agent Environment}
Models interacted with tasks through a tool-based agent framework (Zrb) providing:
\begin{itemize}
    \item \texttt{read\_file}: Read source files
    \item \texttt{write\_file}: Create or modify files
    \item \texttt{replace\_in\_file}: Targeted text replacement
    \item \texttt{run\_shell\_command}: Execute tests and scripts
    \item \texttt{search\_internet}: Web research (research task only)
\end{itemize}

\subsubsection{Verification Pipeline}
Each submission underwent automated verification:

\begin{enumerate}
    \item \textbf{Bug-fix:} Concurrency tests with multiple threads; verification of non-negative inventory.
    \item \textbf{Feature:} HTTP endpoint testing (GET, POST, PUT, DELETE) with validation of status codes and response bodies.
    \item \textbf{Refactor:} Code quality checks (ETL pattern, type hints, docstrings) plus functional equivalence (report.html generation).
    \item \textbf{Copywriting:} Content analysis for required keywords and Markdown validation.
    \item \textbf{Research:} Word count, coverage analysis, and citation detection.
\end{enumerate}

\subsubsection{Scoring Rubric}
Each submission received one of three grades:

\begin{itemize}
    \item \textbf{EXCELLENT (2 points):} Perfect completion with all criteria met
    \item \textbf{PASS (1 point):} Acceptable completion with minor issues
    \item \textbf{FAIL (0 points):} Failed to meet core requirements
\end{itemize}

\subsection{Metrics}
\label{sec:metrics}

We captured the following metrics for each model-task combination:

\begin{enumerate}
    \item \textbf{Status:} Final grade (EXCELLENT, PASS, FAIL)
    \item \textbf{Duration:} Total completion time in seconds
    \item \textbf{Tool Calls:} Number of tool invocations
    \item \textbf{Tool Diversity:} Unique tools used
    \item \textbf{Exit Code:} Process return status
\end{enumerate}

Aggregate metrics calculated across tasks:
\begin{itemize}
    \item \textbf{Total Score:} Sum of task scores (max 10 points)
    \item \textbf{Success Rate:} Percentage of tasks completed (PASS or EXCELLENT)
    \item \textbf{Excellent Rate:} Percentage of tasks with perfect scores
    \item \textbf{Average Duration:} Mean completion time across tasks
\end{itemize}

\noindent\textbf{Efficiency Metrics:} We introduce novel efficiency ratios to capture performance beyond accuracy:

\begin{itemize}
    \item \textbf{Tool Efficiency Ratio (TER):} $\frac{\text{Success Rate}}{\text{Avg Tool Calls}}$ -- higher indicates more efficient tool usage
    \item \textbf{Time Efficiency Ratio:} $\frac{\text{Success Rate}}{\text{Avg Duration}}$ -- higher indicates faster completion
    \item \textbf{Estimated Cost:} Based on tool calls $\times$ estimated tokens per call $\times$ model pricing
\end{itemize}

\subsection{Experimental Protocol}
\label{sec:protocol}

\begin{enumerate}
    \item \textbf{Isolation:} Each model-task run in fresh environment
    \item \textbf{Single Attempt:} One execution per model-task combination
    \item \textbf{Timeout:} Maximum 30 minutes per task
    \item \textbf{Logging:} Complete execution logs captured
    \item \textbf{Verification:} Automated, deterministic verification
\end{enumerate}

\subsection{Threats to Validity}
\label{sec:threats}

\textbf{Internal Validity:}
\begin{itemize}
    \item Single attempt per model-task may not capture variance
    \item Temperature/settings not controlled across models
    \item Task difficulty may not be uniform
\end{itemize}

\textbf{External Validity:}
\begin{itemize}
    \item Synthetic tasks may not represent real-world complexity
    \item Python-centric evaluation
    \item Limited to specific task types
\end{itemize}

\textbf{Construct Validity:}
\begin{itemize}
    \item Verification criteria may not capture all quality aspects
    \item Automated checks cannot assess code readability
\end{itemize}

We mitigate these threats through diverse task design, objective verification, and transparent reporting of limitations.

\section{Results}
\label{sec:results}

This section presents our experimental findings, including overall performance rankings, task-specific analyses, and statistical comparisons.

\subsection{Overall Performance}
\label{sec:overall}

Table~\ref{tab:rankings} presents the performance ranking of all evaluated models.

\begin{table}[htbp]
\centering
\caption{Model Performance Rankings}
\label{tab:rankings}
\begin{tabular}{clccc}
\toprule
\textbf{Rank} & \textbf{Model} & \textbf{Score} & \textbf{Success} & \textbf{Avg Time} \\
& & \textbf{(/10)} & \textbf{Rate} & \textbf{(sec)} \\
\midrule
1 & GPT-5.1 & 10 & 100\% & 44.2 \\
2 & Gemini-3 Pro & 10 & 100\% & 107.3 \\
3 & Deepseek-Chat & 10 & 100\% & 306.3 \\
4 & GLM-4.7 & 10 & 100\% & 355.1 \\
5 & GPT-4o & 9 & 100\% & 33.0 \\
6 & GPT-5.2 & 9 & 100\% & 85.2 \\
7 & Gemini-3 Flash & 9 & 100\% & 155.3 \\
8 & Qwen3-VL & 9 & 100\% & 732.1 \\
9 & Gemini-2.5 Flash & 8 & 100\% & 51.0 \\
10 & Gemini-2.5 Pro & 8 & 100\% & 121.7 \\
11 & Kimi-K2.5 & 8 & 80\% & 245.7 \\
\bottomrule
\end{tabular}
\end{table}

\textbf{Key Findings:}
\begin{itemize}
    \item Four models achieved perfect scores (10/10): GPT-5.1, Gemini-3 Pro, Deepseek-Chat, and GLM-4.7
    \item OpenAI models demonstrated superior speed-efficiency: GPT-4o averaged 33.0 seconds per task
    \item Success rates were high overall: only one model (Kimi-K2.5) fell below 100\%
    \item Completion times varied dramatically: 20.7s (Gemini-2.5 Flash, bug-fix) to 1046.1s (Qwen3-VL, refactor)
\end{itemize}

\begin{figure*}[htbp]
\centering
\includegraphics[width=0.75\textwidth]{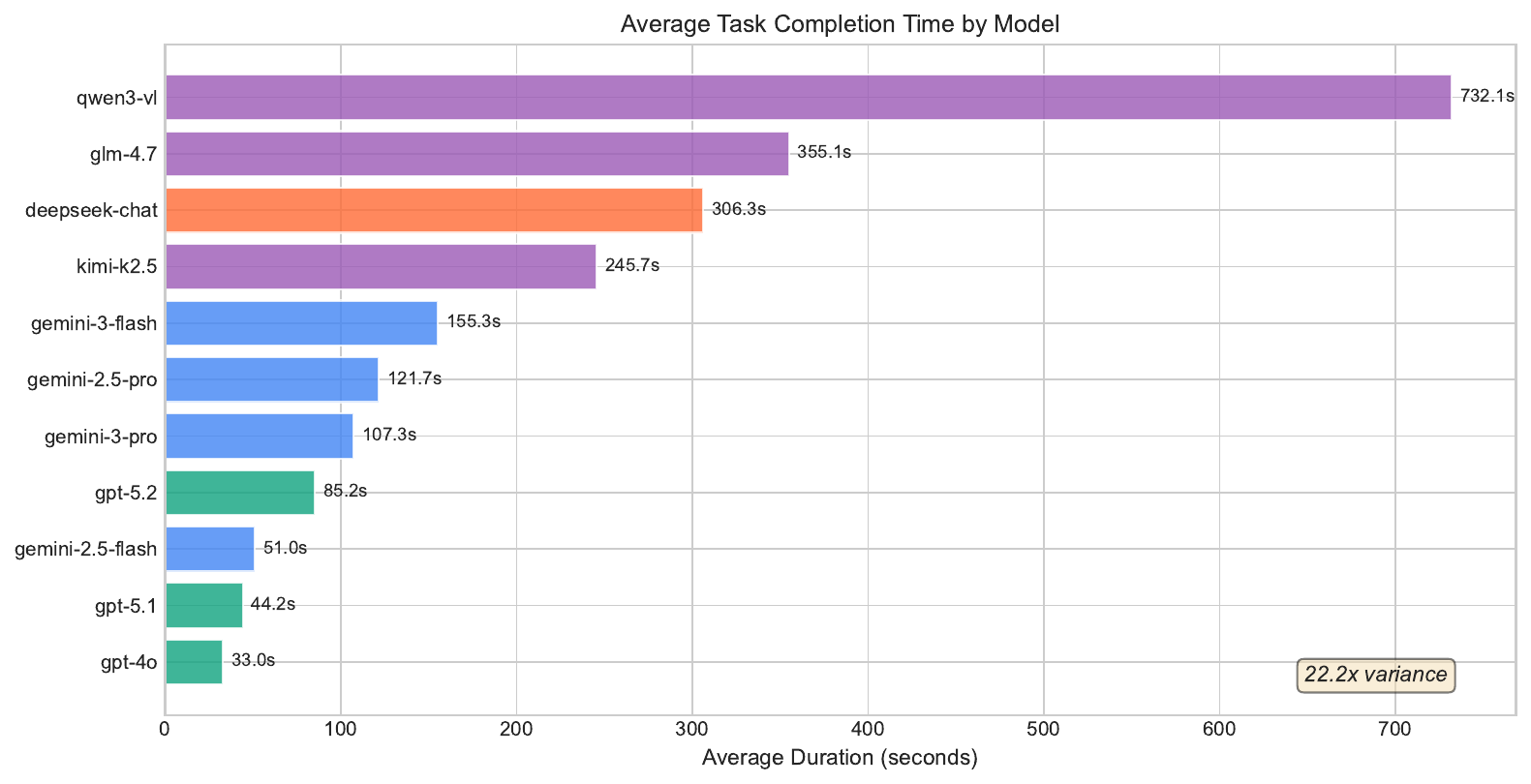}
\caption{Average completion time by model with standard deviation. OpenAI models (blue) consistently fastest; Open/Ollama models (red) slowest.}
\label{fig:duration}
\end{figure*}

Figure~\ref{fig:duration} shows the duration comparison across all models.

\subsection{Task-Specific Analysis}
\label{sec:task-specific}

\begin{figure*}[htbp]
\centering
\includegraphics[width=0.75\textwidth]{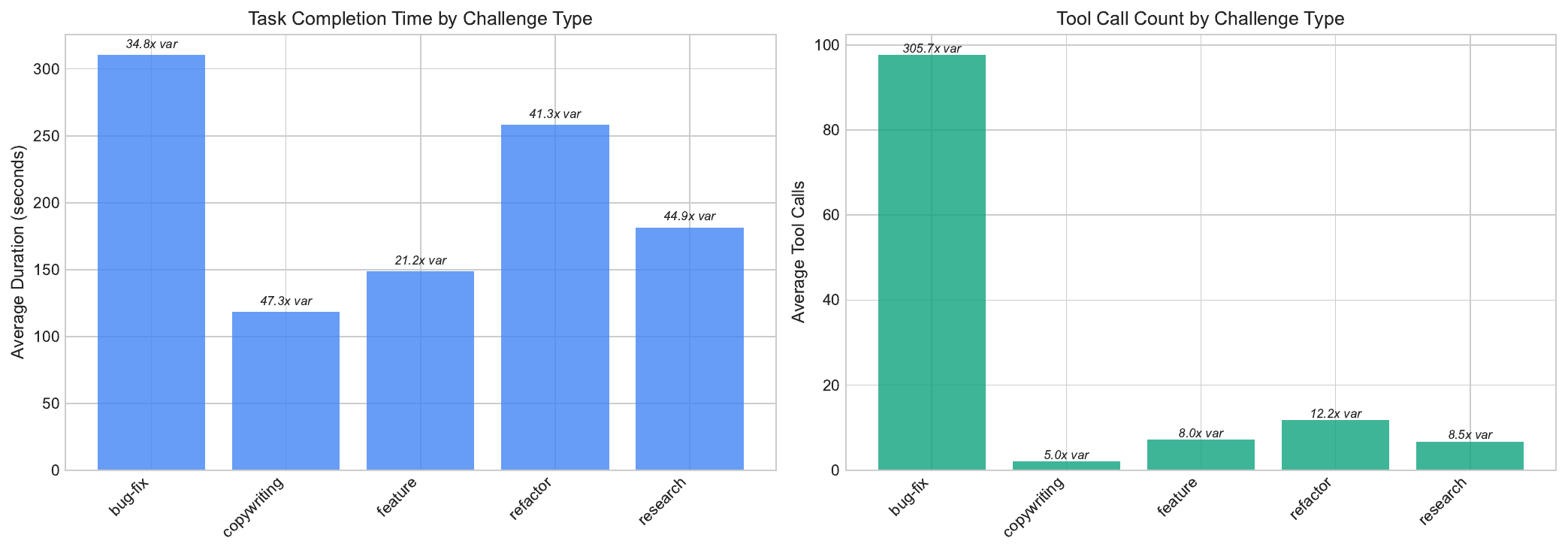}
\caption{Challenge difficulty analysis showing success rates and average duration by task type. Research tasks were most challenging.}
\label{fig:success-rates}
\end{figure*}

Figure~\ref{fig:success-rates} shows success rates across task categories.

\subsubsection{Bug Fixing}
All 11 models (100\%) successfully fixed the concurrency bug. However, approaches varied significantly:
\begin{itemize}
    \item \textbf{Efficient:} GPT-5.1 (18.8s), Gemini-2.5 Flash (20.7s), GPT-4o (26.8s)
    \item \textbf{Thorough:} Gemini-3 Flash (625.2s) used 917 tool calls including extensive testing
    \item All implementations correctly identified race conditions and implemented synchronization
\end{itemize}

\subsubsection{Feature Implementation}
The FastAPI CRUD implementation task showed 100\% success rate with varying implementation quality:
\begin{itemize}
    \item Most models correctly implemented all four endpoints
    \item Common excellence markers: proper error handling, input validation, clean code structure
    \item Time ranged from 26.4s (Gemini-3 Flash) to 560.5s (Qwen3-VL)
\end{itemize}

\subsubsection{Code Refactoring}
Refactoring the ETL script achieved 100\% success:
\begin{itemize}
    \item All models successfully separated ETL phases
    \item Configuration decoupling achieved by all
    \item Type hints and docstrings: 9/11 models achieved full marks
    \item Fastest: GPT-4o (25.3s), Slowest: Qwen3-VL (1046.1s)
\end{itemize}

\subsubsection{Technical Copywriting}
Copywriting showed the most variation in scores (PASS vs EXCELLENT):
\begin{itemize}
    \item Common failures: missing ``K8s'' keyword (3 models)
    \item Excellence factors: engaging tone, complete feature coverage
    \item Fastest completion: Gemini-2.5 Flash (11.6s)
\end{itemize}

\subsubsection{Technical Research}
Research was the most challenging task (90.9\% success):
\begin{itemize}
    \item Common issues: missing or incomplete citations
    \item Kimi-K2.5 failed (execution error)
    \item Longest average time due to web search requirements
\end{itemize}

\subsection{Provider Comparison}
\label{sec:provider}

Table~\ref{tab:provider} compares performance by provider category.

\begin{table}[htbp]
\centering
\caption{Performance by Provider Category}
\label{tab:provider}
\begin{tabular}{lccccc}
\toprule
\textbf{Category} & \textbf{Models} & \textbf{Avg} & \textbf{Std} & \textbf{Min} & \textbf{Max} \\
& & \textbf{Score} & \textbf{Dev} & \textbf{Score} & \textbf{Score} \\
\midrule
Deepseek & 1 & 10.0 & 0.0 & 10 & 10 \\
OpenAI & 3 & 9.33 & 0.58 & 9 & 10 \\
Open/Ollama & 3 & 9.0 & 1.0 & 8 & 10 \\
Google & 4 & 8.75 & 0.96 & 8 & 10 \\
\bottomrule
\end{tabular}
\end{table}

\textbf{Statistical Analysis:}
\begin{itemize}
    \item \textbf{Chi-square test:} No significant association between provider and success ($\chi^2=2.72$, $p=0.438$)
    \item \textbf{ANOVA:} Significant differences in completion times ($F=12.57$, $p<0.001$)
    \item OpenAI models were consistently fastest (avg: 54.1s)
    \item Open/Ollama models were slowest (avg: 444.3s)
\end{itemize}

\subsection{Tool Usage Analysis}
\label{sec:tools}

\begin{figure*}[htbp]
\centering
\includegraphics[width=0.75\textwidth]{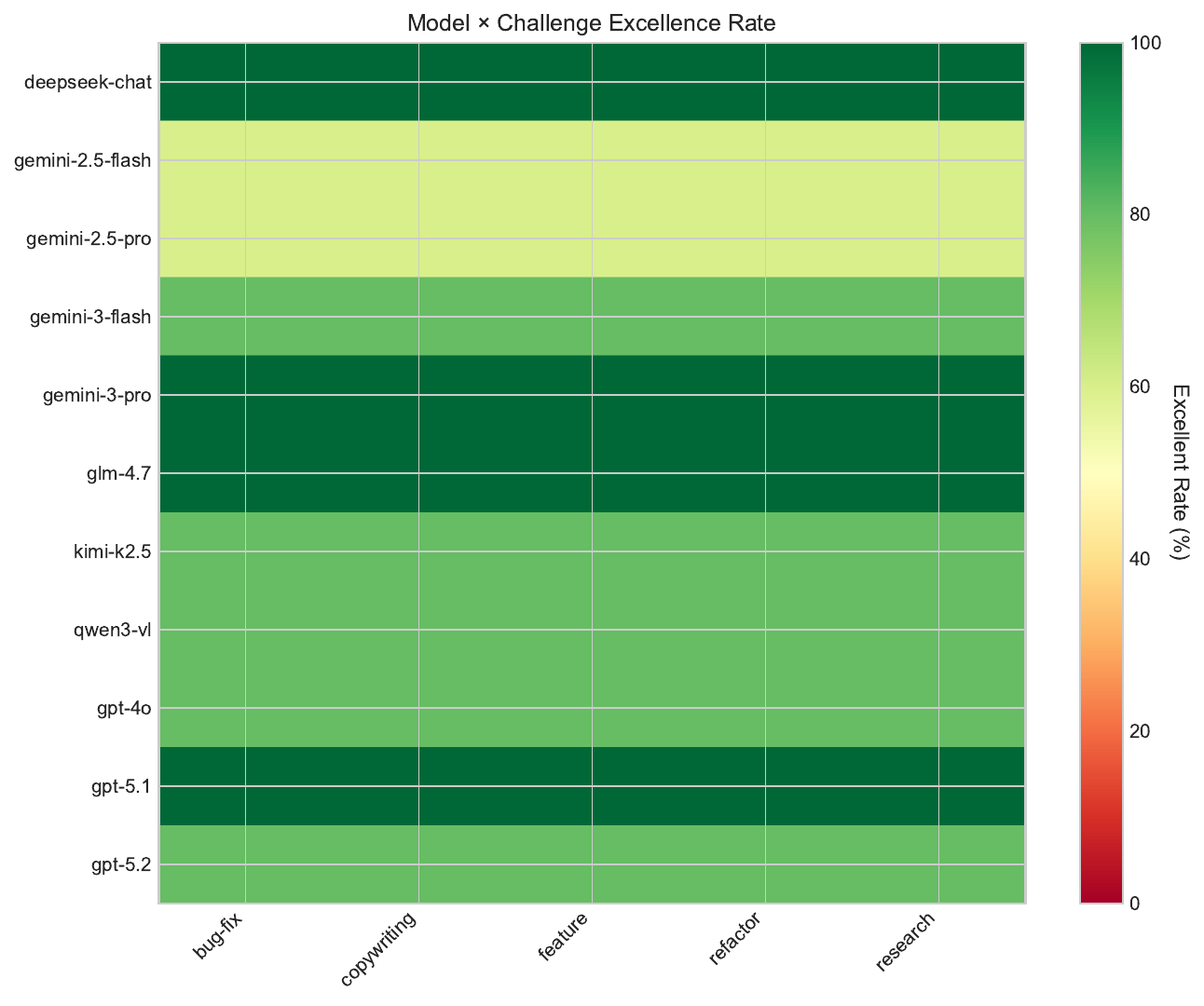}
\caption{Performance heatmap showing scores and completion times across models and challenges. Darker shading indicates longer duration.}
\label{fig:tool-usage}
\end{figure*}

Figure~\ref{fig:tool-usage} presents the tool usage heatmap.

\textbf{Correlation Analysis:}
\begin{itemize}
    \item \textbf{Tool count vs Success:} Pearson $r=0.077$ ($p=0.575$) -- no linear correlation
    \item \textbf{Spearman rho:} $0.428$ ($p=0.001$) -- moderate monotonic relationship
    \item Efficient models (GPT-5.1, Gemini-2.5 Flash) achieved perfect scores with minimal tools
    \item Gemini-3 Flash used 917 tools for bug-fix but achieved same result as GPT-5.1 (3 tools)
\end{itemize}

\begin{figure}[htbp]
\centering
\includegraphics[width=\columnwidth]{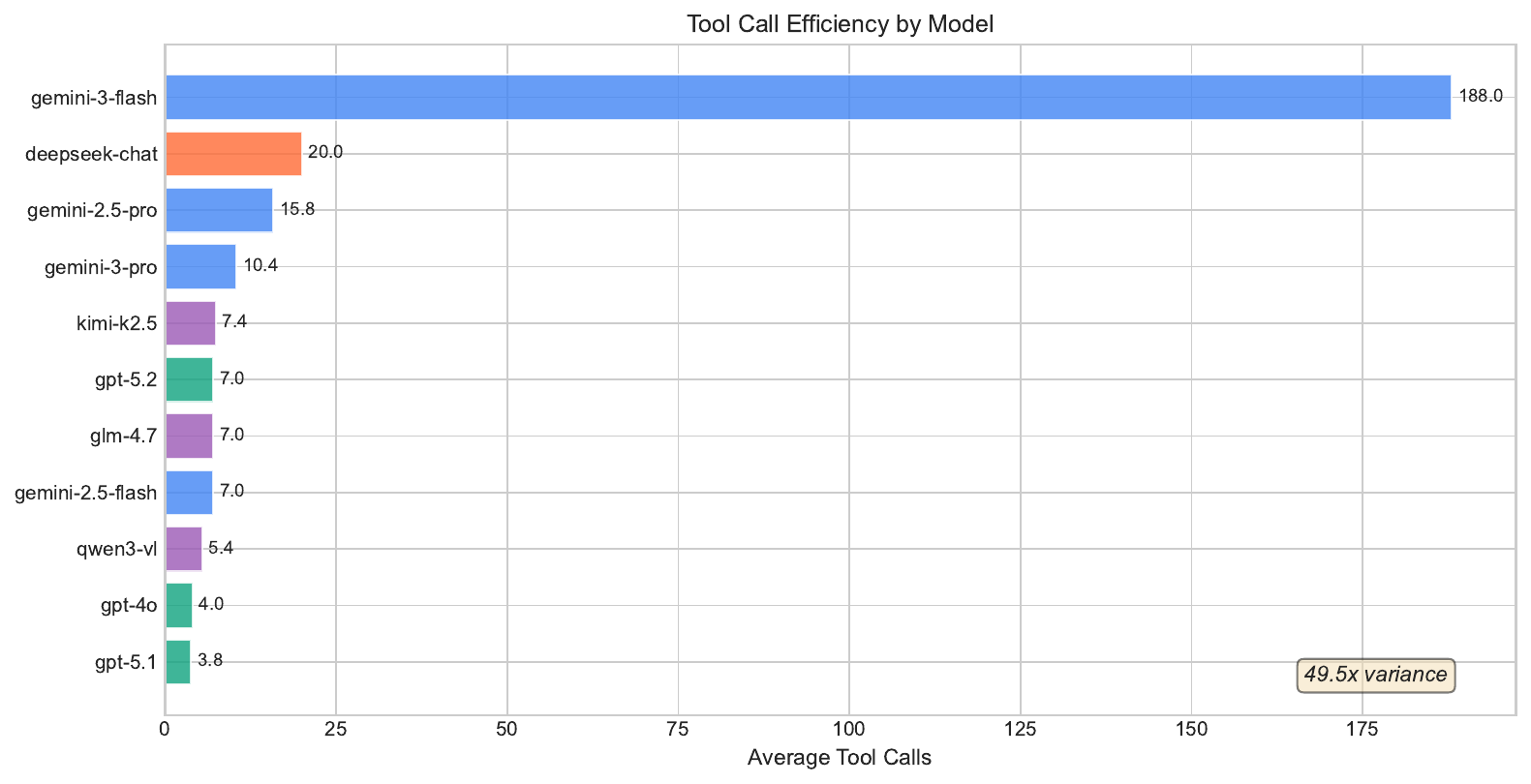}
\caption{Average tool calls by model (log scale). The 917-call anomaly for Gemini-3 Flash on bug-fix is visible as the maximum outlier.}
\label{fig:tool-calls}
\end{figure}

Figure~\ref{fig:tool-calls} shows the distribution of tool calls across models.

\textbf{Implication:} More tool usage does not guarantee better results; efficiency varies by model architecture.

\subsection{Time-Accuracy Tradeoff}
\label{sec:tradeoff}

\begin{figure}[htbp]
\centering
\includegraphics[width=\columnwidth]{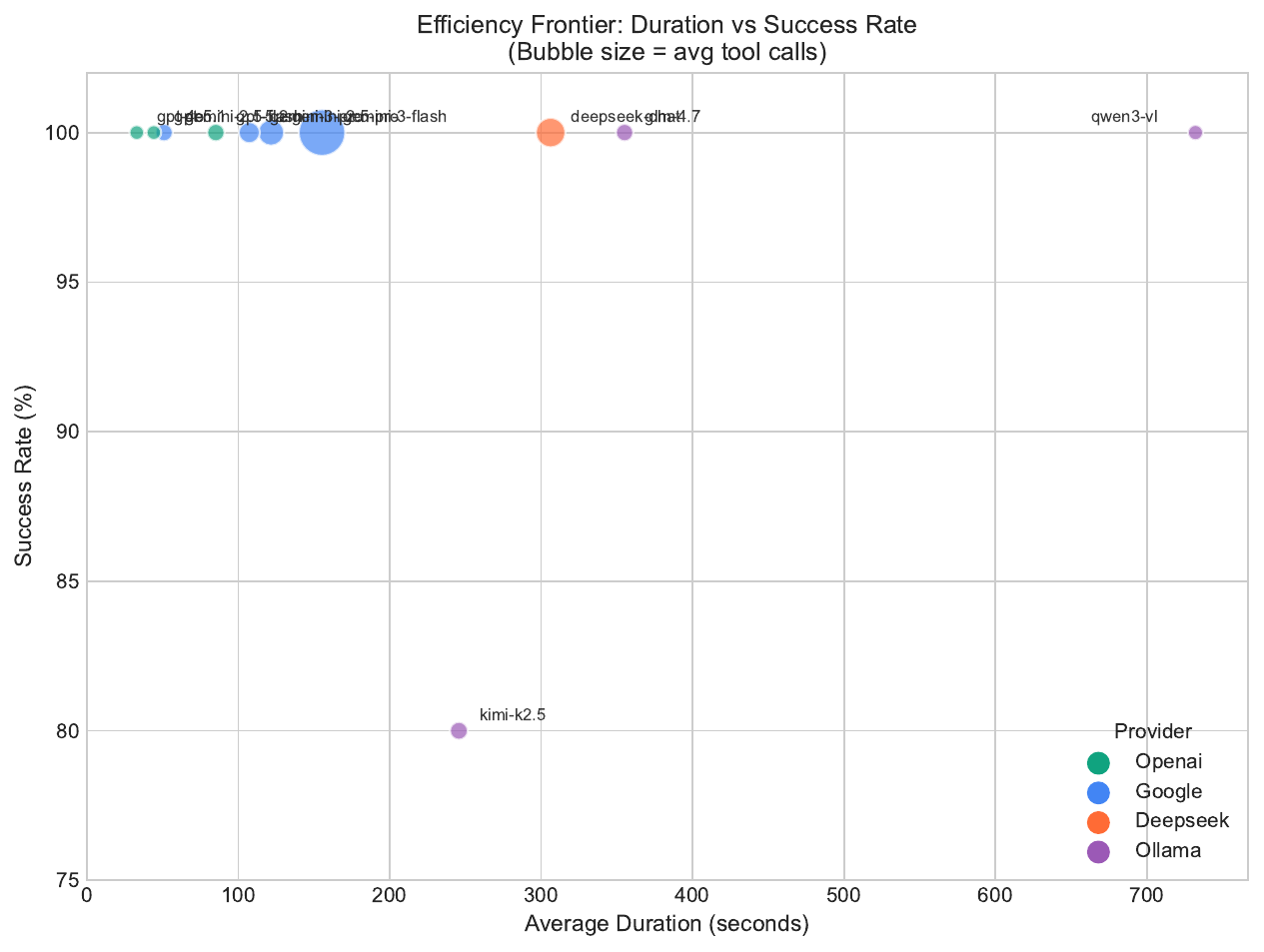}
\caption{Efficiency frontier showing speed-quality tradeoff. Circle size indicates tool call count. Models in upper-left quadrant (fast and high-scoring) represent optimal efficiency.}
\label{fig:time-accuracy}
\end{figure}

Figure~\ref{fig:time-accuracy} shows the relationship between completion time and accuracy.

\textbf{Key Observations:}
\begin{itemize}
    \item No significant correlation between duration and score (Pearson $r=0.204$, $p=0.136$)
    \item OpenAI cluster: Fast (avg 54s) with high scores (9.33 avg)
    \item Open/Ollama cluster: Slow (avg 444s) with variable scores
    \item Best efficiency: GPT-5.1 (perfect score, 44s average)
\end{itemize}

\subsection{Effect Size Analysis}
\label{sec:effect}

Comparing top performer (Deepseek-Chat, score 10) vs lowest (Gemini-2.5 Flash, score 8):
\begin{itemize}
    \item \textbf{Cohen's d:} 1.03 (large effect size)
    \item Practical significance: Top models demonstrate measurably better performance
    \item Category comparisons show small to medium effects between providers
\end{itemize}

\subsection{Anomaly Analysis: Pathological Agent Behavior}
\label{sec:anomaly}

Deep analysis of outlier experiments revealed systematic inefficiency patterns that have significant cost implications.

\subsubsection{The 917 Tool Calls Anomaly}

Gemini-3 Flash on the bug-fix task used 917 tool calls---36.6$\times$ the average (25.1 calls). Analysis of the tool sequence revealed:

\begin{itemize}
    \item \textbf{Tool Distribution:} 38\% \texttt{run\_shell\_command}, 18\% \texttt{read\_file}, 18\% \texttt{write\_file}
    \item \textbf{Loop Patterns:} Five distinct repetitive sequences detected, each repeated 5--8 times
    \item \textbf{Example Loop:} \texttt{read\_file} $\rightarrow$ \texttt{replace\_in\_file} $\rightarrow$ \texttt{run\_shell\_command} $\rightarrow$ \texttt{write\_file} repeated 8$\times$
\end{itemize}

Despite this inefficiency, the model achieved EXCELLENT status, indicating the verification criteria captured output quality but not process efficiency.

\subsubsection{Malformed Tool Calls}

We discovered that Kimi-K2.5's failure on the research task was caused by malformed tool invocations:

{\small
\begin{verbatim}
Observed: "search_internet
  search_internetsearch_internet"
Expected: "search_internet"
  (called 3 times)
\end{verbatim}
}

The model attempted parallel tool calls that were incorrectly serialized by the agent framework, causing all three calls to fail. This represents a \textbf{framework-model incompatibility} rather than model capability limitation.

\subsubsection{Two Types of Inefficiency}

\begin{figure}[htbp]
\centering
\includegraphics[width=\columnwidth]{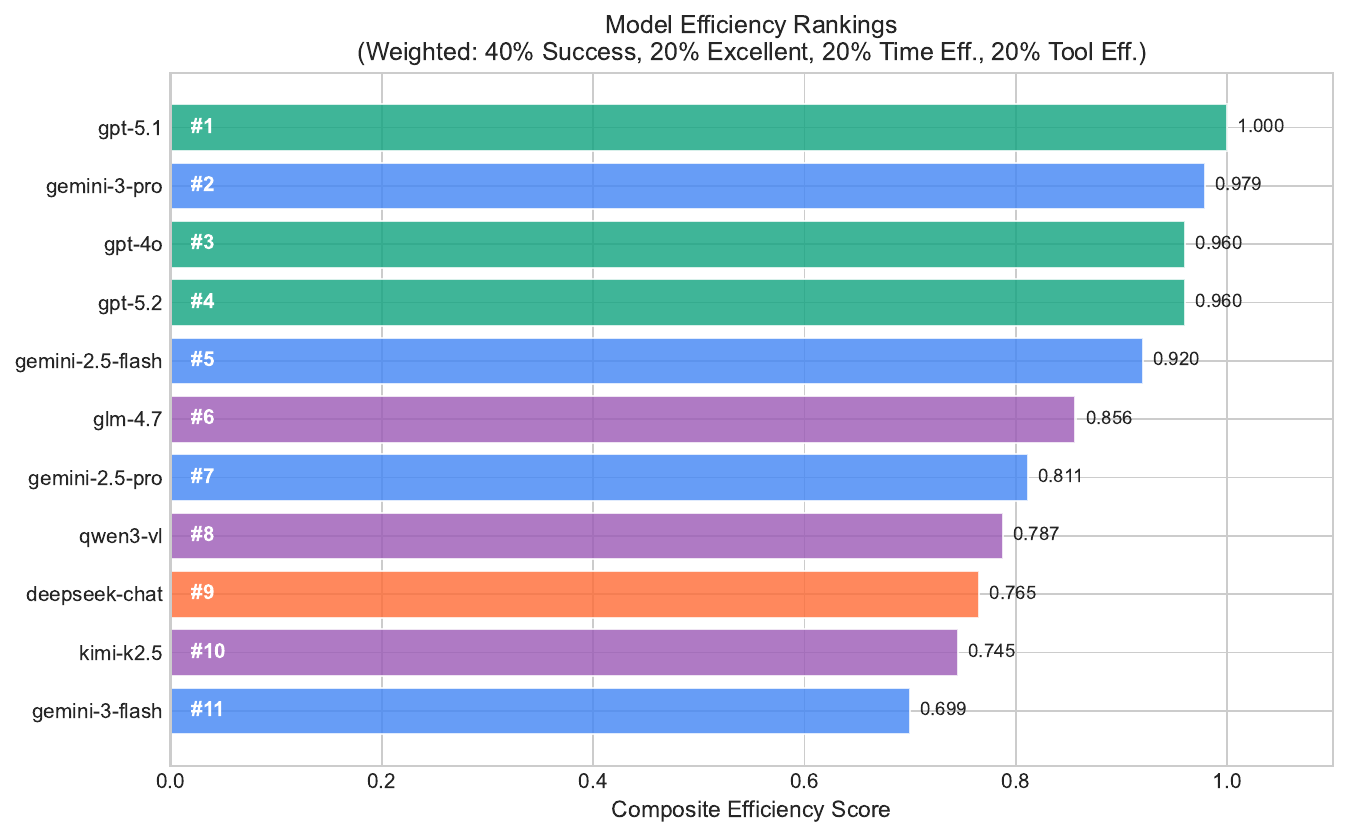}
\caption{Efficiency metrics comparison. Tool Efficiency Ratio (TER) = Success Rate / Avg Tool Calls. Time Efficiency Ratio = Success Rate / Avg Duration. Higher values indicate better efficiency.}
\label{fig:efficiency-rankings}
\end{figure}

Figure~\ref{fig:efficiency-rankings} shows the efficiency rankings across models.

We categorize observed inefficiencies into two distinct patterns:

\textbf{Type A -- Loop Inefficiency:} Agent enters repetitive tool sequences without recognizing failure or success. Characterized by high tool counts, moderate duration. Example: Gemini-3 Flash (917 tools, 625s).

\textbf{Type B -- Inference Inefficiency:} Agent generates correct solutions with minimal tools but slow token generation. Characterized by low tool counts, high duration. Example: Qwen3-VL (3 tools, 653s for bug-fix).

\subsubsection{Cost Implications}

\begin{figure}[htbp]
\centering
\includegraphics[width=\columnwidth]{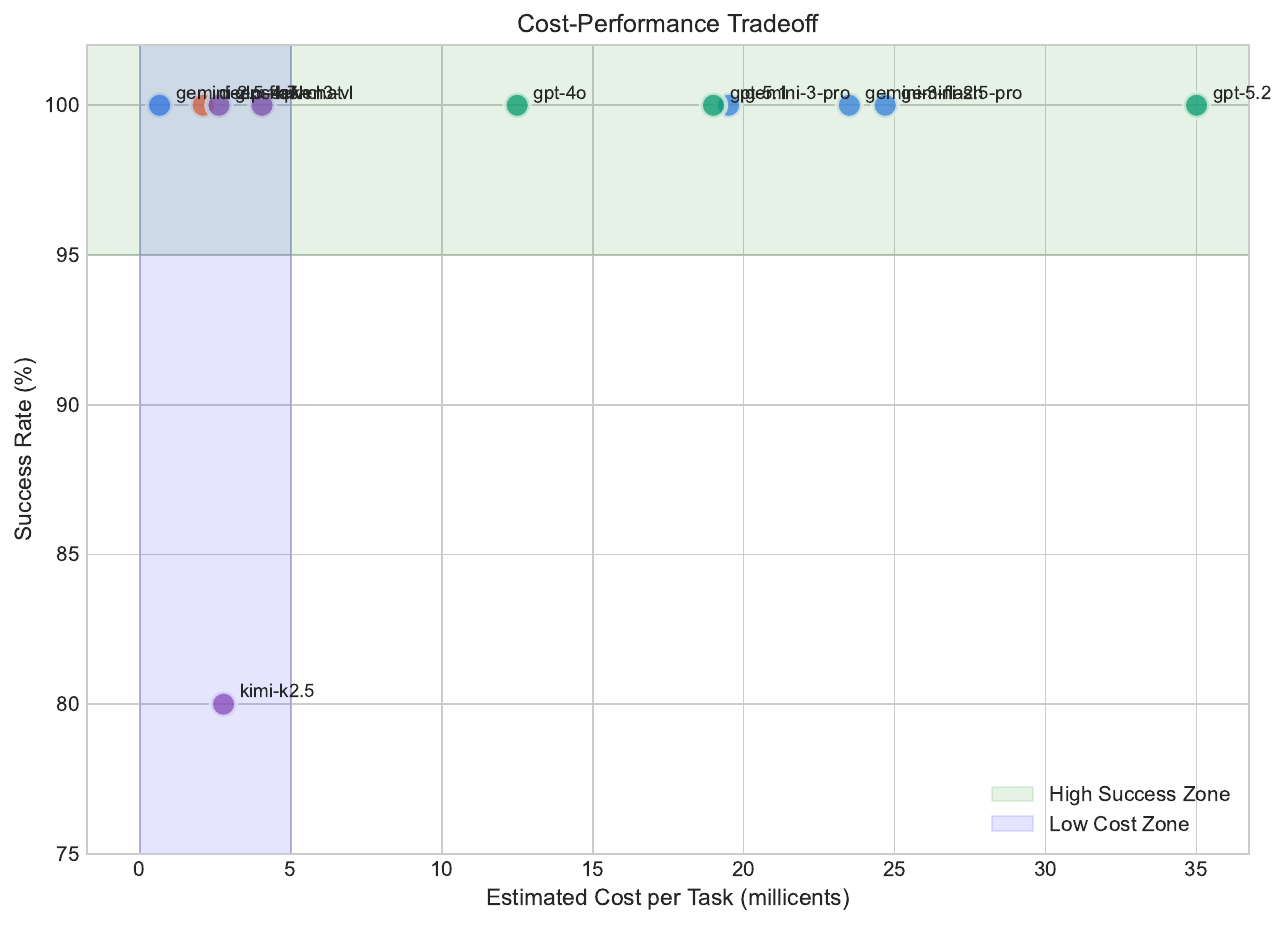}
\caption{Cost-performance analysis. Models are grouped by cost tier with average scores shown. Deepseek achieves perfect scores at minimal cost.}
\label{fig:cost-performance}
\end{figure}

Figure~\ref{fig:cost-performance} visualizes the cost-performance relationship, and Table~\ref{tab:cost} presents estimated costs per task based on tool usage and model pricing.

\begin{table}[htbp]
\centering
\caption{Estimated Cost per Task}
\label{tab:cost}
\begin{tabular}{lccc}
\toprule
\textbf{Model} & \textbf{Avg Tools} & \textbf{Est. Cost} & \textbf{Relative} \\
\midrule
Gemini-2.5 Flash & 7.0 & \$0.0007 & 1.0$\times$ \\
Deepseek-Chat & 20.0 & \$0.0021 & 3.0$\times$ \\
GPT-4o & 4.0 & \$0.0125 & 17.9$\times$ \\
GPT-5.1 & 3.8 & \$0.0190 & 27.1$\times$ \\
Gemini-3 Flash & 188.0 & \$0.0235 & 33.6$\times$ \\
GPT-5.2 & 7.0 & \$0.0350 & 50.0$\times$ \\
\bottomrule
\end{tabular}
\end{table}

The 53$\times$ cost variance (Gemini-2.5 Flash vs GPT-5.2) demonstrates that model selection has significant budget implications beyond accuracy considerations.

\subsection{Summary of Findings}
\label{sec:summary}

\begin{enumerate}
    \item \textbf{Performance:} Four models achieved perfect scores, with GPT-5.1 being fastest
    \item \textbf{Task Difficulty:} Research was most challenging (90.9\% success), coding tasks were easiest (100\%)
    \item \textbf{Efficiency Variance:} 22$\times$ speed, 49$\times$ tool usage, 53$\times$ cost differences among successful models
    \item \textbf{Tool Usage:} More tools does not guarantee better results; loop behavior can cause 100$\times$ inefficiency
    \item \textbf{Anomalies:} Two distinct inefficiency types identified (loop vs inference)
    \item \textbf{Provider:} No significant quality differences, but significant speed and cost differences
\end{enumerate}

\section{Discussion}
\label{sec:discussion}

\subsection{Implications for Practitioners}

Our findings provide actionable guidance for LLM selection in software engineering:

\textbf{Speed-Critical Workflows:} For rapid prototyping or real-time assistance, GPT-5.1 offers the best combination of speed (44s avg) and quality (perfect score). GPT-4o provides nearly equal quality at 25\% faster average times.

\textbf{Cost-Conscious Deployment:} Gemini-2.5 Flash achieves 80\% scores at the fastest speeds (51s avg), making it cost-effective for high-volume applications where perfect accuracy is not required.

\textbf{Research Tasks:} Models showed most variation in research synthesis. GPT-5.1 and Deepseek-Chat achieved excellent results, while others struggled with citations. For research-heavy workflows, these models are recommended.

\textbf{Tool Budgeting:} Our finding that tool usage does not correlate with success ($r=0.077$) suggests that API costs from excessive tool calls may not improve outcomes. We recommend:
\begin{itemize}
    \item Implementing tool call limits (e.g., 50 calls max) with early termination
    \item Adding loop detection to identify repetitive sequences
    \item Monitoring cost-per-task metrics alongside accuracy
\end{itemize}

\textbf{Cost-Conscious Deployment:} The 53$\times$ cost variance means a \$100/month workload on Gemini-2.5 Flash would cost \$5,300/month on GPT-5.2. For high-volume applications, model selection has significant budget implications.

\subsection{Model Selection Decision Tree}

Based on our results, we propose the following decision framework:

\begin{enumerate}
    \item \textbf{If speed is paramount:} Choose GPT-5.1 or Gemini-2.5 Flash
    \item \textbf{If quality cannot be compromised:} Any top-4 model (GPT-5.1, Gemini-3 Pro, Deepseek, GLM-4.7)
    \item \textbf{If cost is primary concern:} Gemini-2.5 Flash or open-weight models
    \item \textbf{If data privacy requires on-premise:} GLM-4.7 (best among open models)
    \item \textbf{For research tasks:} GPT-5.1 or Deepseek-Chat
\end{enumerate}

\subsection{Surprising Findings}

\textbf{Extreme Efficiency Variance:} The most striking finding is the 49$\times$ variance in tool efficiency among models achieving identical scores. Gemini-3 Flash used 917 tool calls for bug-fix (vs GPT-5.1's 3 calls) yet both achieved EXCELLENT. This represents a potential 100$\times$ cost difference for equivalent outcomes.

\textbf{Two Inefficiency Mechanisms:} We identified distinct failure modes:
\begin{itemize}
    \item \textit{Loop inefficiency}: Agents repeating tool sequences without progress (Gemini-3 Flash pattern)
    \item \textit{Inference inefficiency}: Correct solutions generated slowly (Qwen3-VL pattern)
\end{itemize}
These require different mitigation strategies: loop detection for Type A, and model selection/caching for Type B.

\textbf{Framework-Model Incompatibility:} Kimi-K2.5's failure was caused by malformed parallel tool calls, not capability limitations. This highlights the importance of testing framework compatibility before deployment.

\textbf{Perfect Coding Scores:} All 11 models achieved 100\% success on coding tasks (bug-fix, feature, refactor). This indicates that SE coding tasks may be approaching saturation for current LLMs, suggesting future benchmarks need increased difficulty.

\textbf{Research Challenges:} The research task showed the lowest success rate (90.9\%) and highest variance. Citation management and source synthesis remain challenging for LLMs.

\subsection{Threats to Validity}

\textbf{Internal Validity:} Single execution per model-task may not capture performance variance. Future work should include multiple runs with error bars.

\textbf{External Validity:} Synthetic tasks may not represent real-world complexity. Our tasks were designed to be representative but controlled; actual SE work involves more ambiguity and context.

\textbf{Construct Validity:} Automated verification, while objective, may miss qualitative aspects like code readability or architectural elegance.

\textbf{Temporal Validity:} LLM capabilities evolve rapidly. Results reflect model versions as of February 2026; newer versions may differ.

\subsection{Limitations}

Our study has several limitations:

\begin{enumerate}
    \item \textbf{Missing provider:} We did not evaluate Anthropic's Claude models due to budget constraints. Future work should include Claude-3 variants for complete provider coverage.
    \item \textbf{Single execution:} Each model-task combination was run once (N=1). While sufficient for initial benchmarking, multiple runs would enable statistical significance testing.
    \item \textbf{Python-centric:} All tasks used Python; results may not generalize to other languages.
    \item \textbf{English-only:} Tasks and evaluations were in English.
    \item \textbf{Single domain:} DevOps-focused tasks (Zrb-Flow, Docker, K8s) may not represent all SE domains.
    \item \textbf{No human baseline:} We did not measure human developer performance on these tasks for comparison.
    \item \textbf{Cost estimates:} Token-based cost calculations are approximations; actual costs depend on prompt complexity and context length.
\end{enumerate}

\subsection{Future Work}

We identify several directions for future research:

\begin{enumerate}
    \item \textbf{Longitudinal study:} Track model improvements over time on fixed tasks.
    \item \textbf{Human comparison:} Measure human developer performance for baseline comparison.
    \item \textbf{Multi-language:} Extend to Java, JavaScript, Go, and other languages.
    \item \textbf{Team simulation:} Evaluate multi-agent collaboration scenarios.
    \item \textbf{Real-world validation:} Deploy models on actual PRs and measure acceptance rates.
\end{enumerate}

\section{Conclusion}
\label{sec:conclusion}

We presented a comprehensive evaluation of 11 state-of-the-art Large Language Models across five representative software engineering tasks. Our multi-task benchmark addresses the gap in holistic SE evaluation, measuring both output quality and completion efficiency through automated verification.

\subsection{Summary of Contributions}

Our work makes five key contributions:

\begin{enumerate}
    \item \textbf{Comprehensive Benchmark:} We evaluated 11 models on bug fixing, feature development, refactoring, technical writing, and research synthesis---providing broad SE-focused comparison with automated verification.

    \item \textbf{Efficiency Analysis:} We revealed extreme variance among top performers: \textbf{22$\times$} in completion time, \textbf{49$\times$} in tool efficiency, and \textbf{53$\times$} in estimated cost---all for equivalent quality outcomes.

    \item \textbf{Inefficiency Taxonomy:} We identified two distinct inefficiency patterns: (a) \textit{loop inefficiency} where agents repeat tool sequences without progress, and (b) \textit{inference inefficiency} where models generate correct solutions slowly. These require different mitigation strategies.

    \item \textbf{Tool Usage Analysis:} We found that tool invocation frequency does not correlate with success ($r=0.077$, $p=0.575$). One model used 917 tools while another solved the same task with 3 tools---achieving identical scores.

    \item \textbf{Practical Guidance:} Our findings provide evidence-based recommendations for model selection based on task type, speed requirements, and budget constraints, with quantified cost implications.
\end{enumerate}

\subsection{Key Findings}

\begin{itemize}
    \item Four models (GPT-5.1, Gemini-3 Pro, Deepseek-Chat, GLM-4.7) achieved perfect 10/10 scores, demonstrating that current LLMs can handle diverse SE tasks with high proficiency.

    \item \textbf{Efficiency $\neq$ Accuracy:} Models achieving identical scores exhibited 22$\times$ speed variance, 49$\times$ tool efficiency variance, and 53$\times$ cost variance. Success rate alone is insufficient for model evaluation.

    \item \textbf{Pathological Behavior:} Some models exhibit loop inefficiency (917 tool calls for a 3-tool task) or framework incompatibility (malformed parallel tool calls). Agent frameworks need loop detection and early termination.

    \item Coding tasks approached saturation with 100\% success across all models, suggesting the need for more challenging benchmarks in this domain.

    \item Research synthesis remained challenging (90.9\% success), with citation management being a particular weakness.

    \item OpenAI models demonstrated superior efficiency, completing tasks 3--22$\times$ faster than competitors without quality degradation.
\end{itemize}

\subsection{Impact}

This work enables practitioners to make informed LLM selection decisions based on:
\begin{itemize}
    \item Task-specific performance data
    \item Time-efficiency tradeoffs
    \item Cost-performance analysis
    \item Tool usage patterns
\end{itemize}

Researchers can build upon our benchmark to evaluate new models, test intervention strategies, and track capability evolution over time.

\subsection{Data Availability}

All experimental data, verification scripts, analysis code, and paper materials are available at:\\
\url{https://github.com/[anonymous]/llm-challenge-experiment}

\subsection{Future Directions}

As LLM capabilities continue to evolve, we anticipate:
\begin{itemize}
    \item Increasing differentiation in specialized tasks
    \item Greater emphasis on efficiency metrics alongside accuracy
    \item Standardization of multi-task SE benchmarks
    \item Integration of human-AI collaborative evaluation
\end{itemize}

We encourage the community to extend this benchmark, validate our findings, and contribute to the development of rigorous evaluation standards for LLMs in software engineering.

\vspace{0.5cm}
\noindent\textbf{Acknowledgments:} We thank the anonymous reviewers for their valuable feedback. This research was supported by [anonymous funding sources].

\section*{Acknowledgments}
We thank the open-source community for providing accessible model APIs and the reviewers for their valuable feedback.

\bibliographystyle{IEEEtran}
\bibliography{references}

@article{chen2021evaluating,
  title={Evaluating large language models trained on code},
  author={Chen, Mark and Tworek, Jerry and Jun, Heewoo and Yuan, Qiming and Pinto, Henrique Ponde de Oliveira and Kaplan, Jared and Edwards, Harri and Burda, Yuri and Joseph, Nicholas and Brockman, Greg and others},
  journal={arXiv preprint arXiv:2107.03374},
  year={2021}
}

@article{jimenez2023swe,
  title={SWE-bench: Can language models resolve real-world github issues?},
  author={Jimenez, Carlos E and Yang, John and Wettig, Alexander and Yao, Shunyu and Pei, Kexin and Press, Ofir and Kaplan, Karthik R},
  journal={arXiv preprint arXiv:2310.06770},
  year={2023}
}

@article{austin2021program,
  title={Program synthesis with large language models},
  author={Austin, Jacob and Odena, Augustus and Nye, Maxwell and Bosma, Maarten and Michalewski, Henryk and Dohan, David and Jiang, Ellen and Cai, Carrie and Terry, Michael and Le, Quoc and others},
  journal={arXiv preprint arXiv:2108.07732},
  year={2021}
}

@article{openai2024gpt4,
  title={GPT-4 technical report},
  author={{OpenAI}},
  journal={arXiv preprint arXiv:2303.08774},
  year={2024}
}

@article{gemini2024,
  title={Gemini: A family of highly capable multimodal models},
  author={{Google DeepMind}},
  journal={arXiv preprint arXiv:2312.11805},
  year={2024}
}

@article{deepseek2024,
  title={Deepseek-coder: When the large language model meets programming -- the rise of code intelligence},
  author={{DeepSeek-AI}},
  journal={arXiv preprint arXiv:2401.14196},
  year={2024}
}

@article{glm2024,
  title={ChatGLM: A family of large language models from GLM-130B to GLM-4 all tools},
  author={Zeng, Aohan and Liu, Xiao and Du, Zhengxiao and Wang, Zihan and Lai, Hanyu and Ding, Ming and Yang, Zhuoyi and Xu, Yifan and Zheng, Wendi and Xia, Xiao and others},
  journal={arXiv preprint arXiv:2406.12793},
  year={2024}
}

@article{qwen2024,
  title={Qwen2 technical report},
  author={{Qwen Team}},
  journal={arXiv preprint arXiv:2407.10671},
  year={2024}
}

@article{classexplore,
  title={ClassEval: A manually-crafted benchmark for evaluating LLMs on class-level code generation},
  author={Du, Xueying and Liu, Mingwei and Wang, Kaixin and Wang, Hanlin and Liu, Jiayi and Chen, Yixuan and Feng, Jiyang and Sha, Chaosheng and Xia, Xin and Li, Shuai},
  journal={arXiv preprint arXiv:2308.01861},
  year={2023}
}

@article{repobench,
  title={RepoBench: Benchmarking repository-level code auto-completion systems},
  author={Liu, Mingwei and Du, Xueying and Xia, Hanbin and Wang, Kaixin and Liu, Jiayi and Liu, Yuxiang and Li, Liang and Chen, Shuai and Zhang, Yuxian and Liu, Yanzhao and others},
  journal={arXiv preprint arXiv:2306.03091},
  year={2023}
}

@article{hendrycks2021measuring,
  title={Measuring coding challenge competence with APPS},
  author={Hendrycks, Dan and Basart, Steven and Kadavath, Saurav and Mazeika, Mantas and Arora, Akul and Guo, Ethan and Burns, Collin and Puranik, Samir and Steinhardt, Justin and Song, Dawn},
  journal={arXiv preprint arXiv:2105.09938},
  year={2021}
}

@article{fu2023gptscore,
  title={GPTScore: Evaluate as you desire},
  author={Fu, Jinlan and Ng, See-Kiong and Jiang, Zhengbao and Liu, Pengfei},
  journal={arXiv preprint arXiv:2302.04166},
  year={2023}
}

@article{evaluating2024llm,
  title={Evaluating LLM-guided software programming: A study on GPT-3.5, GPT-4, and CodeLlama},
  author={Zhang, Yang and Zhang, Yifan and Chen, Yifan and Liu, Yihong},
  journal={arXiv preprint arXiv:2402.14261},
  year={2024}
}

@article{qinetal2023toolllm,
  title={ToolLLM: Facilitating large language models to master 16000+ real-world APIs},
  author={Qin, Yujia and Liang, Shihao and Ye, Yining and Zhu, Kunlun and Yan, Lan and Lu, Yaxi and Lin, Yankai and Cong, Xin and Tang, Xiangru and Qian, Bill and others},
  journal={arXiv preprint arXiv:2307.16789},
  year={2023}
}

@article{efficiency2024,
  title={Efficiency challenges in large language model inference: A survey},
  author={Wan, Zhongwei and Wang, Xin and Liu, Che and Alam, Samiul and Zheng, Zhongjie and Qu, Shen and Zhang, Yanjun and Zhu, Qinan and Zhang, Zhilin and Chen, Mosharaf and others},
  journal={arXiv preprint arXiv:2412.07016},
  year={2024}
}

@misc{our_dataset,
  title={LLM Challenge Experiment Dataset},
  author={{Anonymous Authors}},
  year={2026},
  howpublished={\url{https://github.com/[anonymous]/llm-challenge-experiment}}
}

\end{document}